\pdfoutput=1 
\documentclass[aps,prl,twocolumn,superscriptaddress,amsfont,graphicx,nofootinbib,preprintnumbers]{revtex4-2}
\usepackage{color,graphicx,epsfig}
\usepackage{ifpdf}
\usepackage{amsmath}
\usepackage{bm}
\usepackage{color}
\usepackage[english]{babel}
\usepackage{graphicx}
\usepackage{amsfonts}
\usepackage{amssymb}
\usepackage{braket}
\usepackage{hyperref}
\usepackage{xspace}
\usepackage{enumerate}

\definecolor{nicered}{rgb}{0.7,0.1,0.1}
\definecolor{nicegreen}{rgb}{0.1,0.5,0.1}
\hypersetup{colorlinks,citecolor= nicegreen,linkcolor= nicered}

\def\beq{\begin{equation}}
\def\eeq{\end{equation}}
\def\bea{\begin{eqnarray}}
\def\eea{\end{eqnarray}}
\def\bsp#1\esp{\begin{split}#1\end{split}}

\newcommand{\eps}{\epsilon}
\newcommand{\ftilde}{\tilde{f}}
\newcommand{\tf}{\ftilde}
\newcommand{\ord}{\mathcal{O}}

\DeclareMathOperator{\Li}{Li}

\newcommand{\cN}{\mathcal{N}}
\newcommand{\cL}{\mathcal{L}}

\newcommand{\figref}[1]{fig.~\ref{#1}}
\newcommand{\zbar}{\bar{z}}
\DeclareMathOperator{\C}{{\bf C}}

\DeclareMathOperator{\Sa}{{\bf \hat{S}}}
\DeclareMathOperator{\deltabar}{\bar{\delta}}
\DeclareMathOperator{\Disc}{Disc}
\DeclareMathOperator{\Tr}{Tr}

\newcommand{\zb}{\bar{z}}

\begin{document}

\preprint{BONN-TH-2025-02}

\title{Antipodal self-duality of square fishnet graphs}

\author{Lance~J.~Dixon}
\affiliation{SLAC National Accelerator Laboratory, Stanford University
Stanford, CA 94309, USA}

\author{Claude~Duhr}
\affiliation{Bethe Center for Theoretical Physics, Universität Bonn, D-53115, Germany}

\date{\today}

\preprint{}

\begin{abstract}
In strongly-deformed planar $\cN=4$ super-Yang-Mills theory, or fishnet theory, a point-split single-trace correlation function of four dimension-$m$ scalar operators is given by a single Feynman integral, which involves integrating over locations of a $m\times m$ grid of points.  We show that for any integer $m$ this square fishnet graph is invariant under the combined action of a kinematic map and the antipode map of the Hopf algebra on multiple polylogarithms, i.e.~it possesses an antipodal self-duality.  
\end{abstract}

\maketitle
%%%%%%%%%%%%%%%%%%%%%%%%%%

Scattering amplitudes, form factors, and correlation functions may exhibit symmetries of a quantum field theory (QFT) that are not manifest from its off-shell Lagrangian formulation. Some of the most prominent examples of hidden symmetries arise in the planar limit of $\cN=4$ maximally supersymmetric Yang-Mills (SYM) theory. Besides superconformal symmetry, on-shell scattering amplitudes in planar $\cN=4$ SYM also enjoy a \emph{dual} superconformal symmetry~\cite{Drummond:2006rz,Bern:2006ew,Alday:2007hr,Bern:2007ct,Drummond:2007cf,Drummond:2007au,Drummond:2008vq}, which closes with ordinary conformal symmetry to form an infinite-dimensional Yangian algebra~\cite{Drummond:2009fd}. 

Recently a novel symmetry was observed for certain form factors and scattering amplitudes in planar $\cN=4$ SYM. Typical symmetries or dualities act on the states of the theory and/or the (dual) coordinates but leave the functional form of observables alone. In contrast, this new symmetry changes observables in a number-theoretic way. To describe its action, we need to recall that perturbative scattering amplitudes and form factors often evaluate to multiple polylogarithms (MPLs)~\cite{chen1977iterated,Goncharov:2001iea,Goncharov:2010jf,Duhr:2011zq,Duhr:2012fh}. MPLs (or rather, their motivic avatars) are endowed with a lot of mathematical structure. In particular, they can be equipped with a coaction~\cite{brownmixedZ}, which essentially decomposes MPLs into simpler MPLs.  Regarded modulo their branch cuts, MPLs form a Hopf algebra~\cite{Goncharov:2005sla}. The maximal iteration of the coaction is called the \emph{symbol}~\cite{Goncharov:2010jf}, and it allows one to represent MPLs in terms of \emph{words} drawn from \emph{letters} belonging to an \emph{alphabet}.  The Hopf algebra contains a ``coinverse" called the antipode $S$, which reverses the order of all the letters in any term in the symbol. (It is also defined on the full Hopf algebra, not just the symbol.)  The Hopf algebra and symbol have proven very powerful for manipulating MPLs~\cite{Duhr:2012fh}, in perturbative computations in the Standard Model as well as in planar $\cN=4$ SYM.  In the latter theory, the amplitude bootstrap program relies on these mathematical structures; it has led to the determination of several amplitudes and form factors to up to eight loops~\cite{Caron-Huot:2020bkp,Dixon:2022rse,Dixon:2023kop,Basso:2024hlx}.

 The new symmetry is called \emph{antipodal self-duality}. It is obtained by acting with $S$ on an observable, followed by a suitable \emph{kinematic map} on the kinematic variables.  The maximally helicity-violating (MHV) four-particle form factor for the chiral part of the stress-tensor supermultiplet in planar $\cN = 4$ SYM (``${\rm Tr}\,\phi^2$'') is invariant under such an antipodal self-duality through three loops at symbol level~\cite{Dixon:2022xqh} and two loops at function level~\cite{Dixon:2024yvq}, when the four particle momenta are constrained to three dimensions. Moreover, an antipodal duality maps the MHV six-particle scattering amplitude to the MHV three-particle form factor for ${\rm Tr}\,\phi^2$~\cite{Dixon:2021tdw}.  Because these latter two quantities are just different limits of the four-particle ${\rm Tr}\,\phi^2$ form factor, its self-duality actually implies the latter duality.  On the other hand, the latter quantities are simpler and can be computed to eight loops, providing a stringent test~\cite{Dixon:2022rse,Dixon:2023kop}.
While these observations establish antipodal (self-)duality for certain MHV form factors and amplitudes in planar $\cN=4$ SYM, its existence remains conjectural at higher loops. Also, its physical origin is completely mysterious.  It would be interesting to find other quantities that exhibit an antipodal duality, that could perhaps shed light on why the symmetry exists.

It has been observed~\cite{Arkani-Hamed:2017ahv} that the antipode action on one-loop integrals is equivalent to inverting the Cayley matrix that multiplies the Feynman parameters. This kinematic map generically maps massless internal lines to massive ones, and so its interpretation in the context of theories with fixed particle mass is uncertain.

\begin{figure}
    \includegraphics[width=0.4\textwidth]{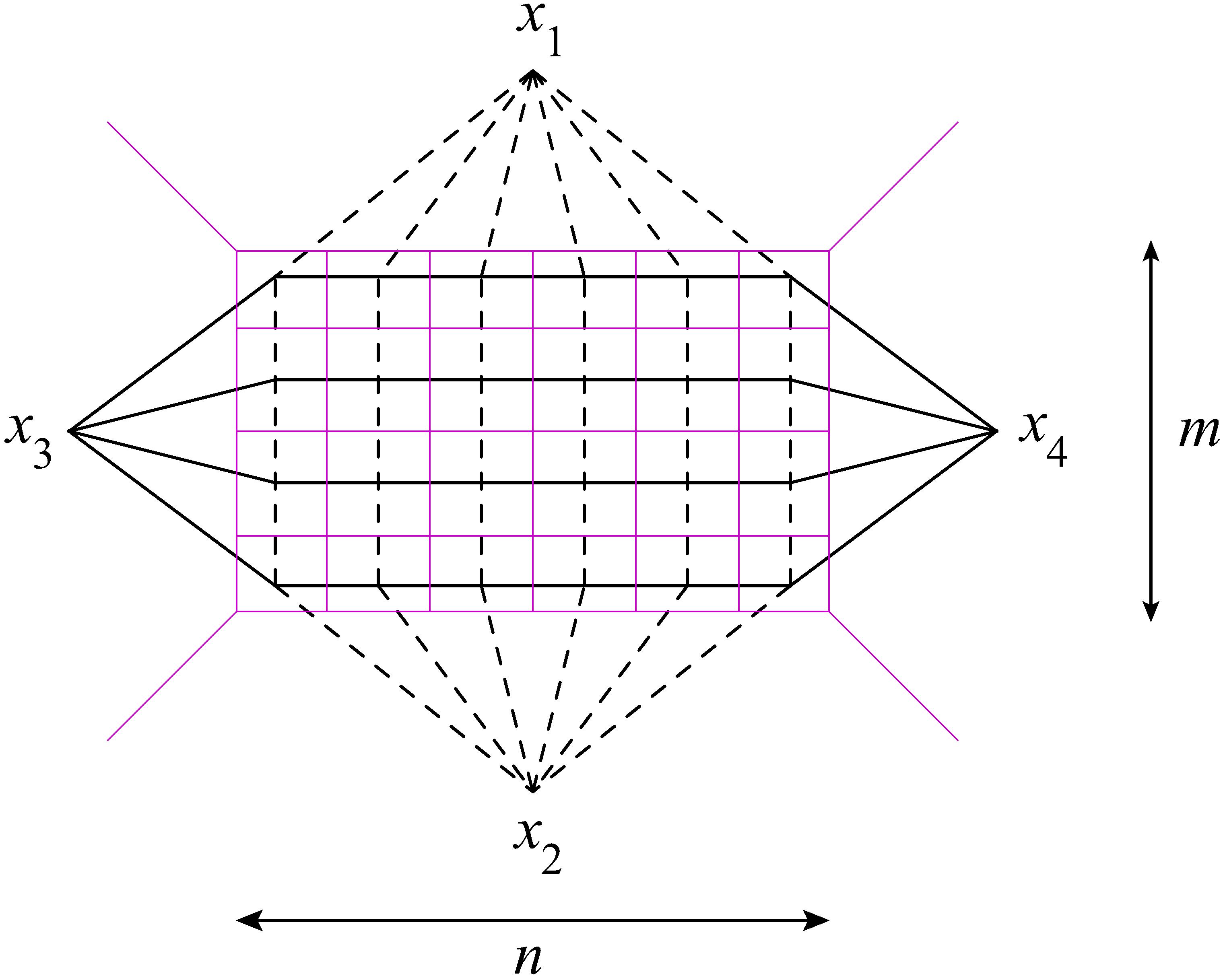}
  \caption{The rectangular fishnet graph for the four-point correlation function $\langle \Tr[ X^n\,Z^m\bar{X}^n\,\bar{Z}^m ] \rangle$, or $G_{m,n}$. The pink lines provide a dual, scattering interpretation.}
  \label{fig:nmfish}
\end{figure}

The goal of this paper is to show that certain correlation functions in the conformal fishnet theory~\cite{Gurdogan:2015csr,Caetano:2016ydc} in four dimensions enjoy antipodal self-duality. Fishnet theory is a deformation of planar $\cN=4$ SYM theory. Its spectrum only contains two complex adjoint scalars $X$ and $Z$, which interact via the non-hermitian Lagrangian
\beq
\mathcal{L} = \Tr\Big[X(-\square)\bar{X} + Z(-\square)\bar{Z} - (4\pi g)^2\,XZ\bar{X}\bar{Z}\Big]\,,
\eeq
where $N_c$ is the number of colors and $g^2$ is the ('t Hooft) coupling constant.
The four-point correlator $\langle \Tr [ X^n(x_1) Z^m(x_3) \bar{X}^n(x_2) \bar{Z}^m(x_4) ] \rangle$ is given by a single Feynman integral $G_{m,n}$, called a fishnet integral (see~fig.~\ref{fig:nmfish}). We will show that square fishnet integrals with $n=m$ are antipodally self-dual. Unlike previous observations of antipodal dualities, we can provide a proof for all ``loop orders'', i.e.~all values of $m$. Moreover, the antipodal symmetry can be realized fully at the function level, and we show how the kinematic map on symbol letters arises in this case. 

%%%%%%%%%%%%%%%%%%%%%%%%%%%%%%%%%%%%%%%%%%
%%%%%%%%%%%%%%%%%%%%%%%%%%%%%%%%%%%%%%%%%%

Our letter is organized as follows: We first provide some background on fishnet integrals and the antipode. We then define the antipodal symmetry which leaves square fishnet integrals invariant and we present a complete proof of the antipodal self-duality. Finally we draw our conclusions. We include two appendices where we collect technical details of the proof omitted in the main text.

%%%%%%%%%%%%%%%%%%%%%%%%%%%%%%

\section{Fishnet integrals}
\label{sec:fishnet}

We focus on \emph{fishnet four-point integrals} in four dimensions, which are defined as the Feynman integrals $G_{m,n}$ in position space depicted in~\figref{fig:nmfish}, with four fixed external points $x_I$, $I=1,2,3,4$. The solid and dashed black lines depict massless propagators of the form $1/x_{ij}^2$ between points $x_i$ and $x_j$, and the $nm$ integration points are arranged to form a $n\times m$ rectangle. One can show that fishnet integrals are conformally invariant~\cite{Drummond:2006rz}. Up to an overall factor that captures the conformal weights, they only depend on two conformal cross ratios,
\beq
G_{m,n} = \frac{g^{2mn}}{x_{12}^{2n} x_{34}^{2m}} \biggl[ \frac{(1-z)(1-\zbar)}{z-\zbar} \biggr]^m \frac{\phi_{m,n}(z,\bar{z})}{{\cal N}}\,,
\eeq
with $x_{ij} = x_i-x_j$, the variables $(z,\zbar)$ are defined by
\beq
\frac{x_{14}^2x_{23}^2}{x_{12}^2x_{34}^2}=\frac{z\zbar}{(1-z)(1-\zbar)}\,,\quad \frac{x_{13}^2x_{24}^2}{x_{12}^2x_{34}^2}=\frac{1}{(1-z)(1-\zbar)}\,,
\eeq
and the normalization factor is
\beq 
{\cal N} = \prod_{k=0}^{2m-1} (n-m+k)! \,.
\eeq
 It was conjectured~\cite{Basso:2017jwq}, and later proven~\cite{Derkachov:2019tzo,Derkachov:2020zvv,Basso:2021omx}, that fishnet integrals can be written as determinants,
\beq\bsp\label{eq:BD}
\phi_{m,n}&\, = \det(f_{n-m+i+j-1})_{1\le i,j\le m}\,,
\esp\eeq
where the arguments of the determinant are \emph{ladder integrals}~\cite{Usyukina:1992jd,Usyukina:1993ch}
\begin{align}\label{eq:ladder}
&f_p(z,\zbar)
= \sum_{j=p}^{2p}\frac{(p-1)!j!\ \cL^{2p-j}}{(j-p)!(2p-j)!}\,\big(\Li_{j}(z)-\Li_j(\zbar)\big)\,,
\end{align}
and ${\cal L} \equiv-\log(z\zbar)$.  Note from~\eqref{eq:BD} that $f_p= \phi_{1,p}$.

In the remainder of this paper we will mostly be interested in square fishnet integrals $\phi_{m,m}$.
In the region where $z$ and $\zbar$ are complex conjugates of each other, ladder and fishnet integrals can be written as single-valued (real analytic) combinations of holomorphic and antiholomorphic logarithms and classical polylogarithms $\Li_n(z) = \sum_{k=1}^\infty \frac{x^k}{k^{n}}$~\cite{Ramakrishnan:1986,Wojtkowiak:1989,Zagier:1991}. 

It is easy to compute the action of the antipode on logarithms and classical polylogarithms. We have (cf.,~e.g.,~\cite{Goncharov:2005sla}),
\beq\bsp\label{eq:Li_antipode}
S(\log z) &\,= -\log z\,,\\
 S\big(\Li_n(z)\big) &\,= \sum_{k=0}^{n-1}\frac{(-1)^{k+1}}{k!}\,\log^k z\,\Li_{n-k}(z)\,.
\esp\eeq
The antipode is linear, involutive ($S^2=\mathrm{id}$) and respects products $S(a\cdot b) = S(a)S(b)$.
Note that, strictly speaking, the antipode acts on the de Rham analogues of (poly)logarithms, which can be understood as (motivic) polylogarithms modulo $i \pi$ (see, e.g.,~\cite{brownnotesmot}). In the following, the distinction will not be crucial, and we will not dwell on it further.

%%%%%%%%%%%%%%%%%%%%%%%%%%%%%

\section{Antipodal self-duality of fishnets}
\label{sec:ASD}
While the combination $\phi_{m,n}$ of polylogarithms in~\eqref{eq:ladder} is single-valued, in general $S(\phi_{m,n})$ is not. The antipode by itself cannot be a symmetry of fishnet graphs. Instead we consider the twisted antipode $\Sa = \C\! S$, where $\C$ is linear and acts on products of holomorphic and antiholomorphic hyperlogarithms $H_p(z)$ and $H_q(\zbar)$ (e.g., classical or harmonic polylogarithms~\cite{Remiddi:1999ew}) of weights $p$ and $q$ as
\beq\label{eq:C_def}
\C\!\big(H_p(z)\,H_q(\zbar)\big) = (-1)^{q}\,H_p(\zbar)\,H_q(z)\,.
\eeq
It is easy to check that $\C$ and $\Sa$ are involutive on functions of even weight $p+q$ and respect products. 

It is instructive to understand how the map $\Sa$ acts at the level of the symbol. The symbol letters of a holomorphic classical or harmonic polylogarithm are of the form $L_i\in\{z,\bar{z},1-z,1-\bar{z}\}$.
Then $\Sa$ acts on symbols of products of holomorphic and antiholomorphic hyperlogarithms via
\beq
\Sa(L_{i_1}\otimes \cdots \otimes L_{i_k}) = (-1)^k\,\mu(L_{i_k})\otimes\cdots\otimes\mu(L_{i_1})\,,
\eeq
where, from \eqref{eq:C_def}, the map $\mu$ conjugates the holomorphic letters and both conjugates and inverts the antiholomorphic letters,
\beq\bsp\label{eq:map_z}
\mu(z) &= \zbar,\quad \mu(1-z)=1-\zbar, \\ \mu(\zbar)&= \frac{1}{z},\quad \mu(1-\zbar)=\frac{1}{1-z}\,,
\esp\eeq
The simplest case of $p=1$ is illustrative.  From \eqref{eq:ladder},
\beq
f_1 = \log(z\zbar) \log\Bigl(\frac{1-z}{1-\zbar}\Bigr) + 2(\Li_2(z)-\Li_2(\zbar))\,,
\label{eq:peq1}
\eeq
whose symbol is
\beq
z\zbar \otimes \frac{1-z}{1-\zbar} - (1-z)(1-\zbar) \otimes \frac{z}{\zbar} \,.
\label{eq:peq1symb}
\eeq
We see that the two terms in \eqref{eq:peq1symb} exchange under $\Sa$.

More generally, at the level of the symbol, $\Sa$ acts via the antipode, followed by the kinematic map $\mu$ on the symbol letters. The map $\Sa$ has exactly the same structure as the antipodal map for amplitudes and form factors in~\cite{Dixon:2021tdw,Dixon:2022xqh,Liu:2022vck}. In our case, however, we can prove antipodal symmetry at function-level for square fishnets with arbitrary $m$.  (The antipodal action can be defined beyond symbol-level for amplitudes and form factors, but it is difficult to check it explicitly beyond four loops, except at specific points where it can be checked in some cases through 8 loops~\cite{Dixon:2021tdw,Dixon:2022rse,Dixon:2023kop}.) 

The main result of this paper is that square fishnet integrals $\phi_{m} \equiv \phi_{m,m}$ are antipodally self-dual, i.e., they are invariant under the antipodal map $\Sa$,
\beq\label{eq:square_ASD}
\Sa\big(\phi_{m}\big) = \phi_{m}\,.
\eeq
The proof, which will be provided in the next section, proceeds in two steps. First, we show that $\Sa$ acts on ladder integrals via
\beq
\Sa\big(f_p\big) = \tilde{f}_p \equiv \sum_{k=0}^{p-1}\binom{p-1}{k}\,L^k\,f_{p-k}\,,
\label{flipf}
\eeq
with $L \equiv \log z\,\log\zbar$. The ladder integrals $f_p$ are single-valued, but the function $L$ is not. Hence, $\Sa\big(f_p\big)$ is not single-valued for $p>1$, and so the ladder integrals $f_p$ and the fishnet integrals $\phi_{m,n}$ are in general not antipodally self-dual. For $m=1$, we see from~\eqref{flipf} (and from the example) that $f_1=\phi_1$ is antipodally self-dual, $\Sa(f_1) = f_1$. In a second step, we then show that~\eqref{flipf}, together with the determinantal structure of square fishnet integrals, leads to antipodal self-duality~\eqref{eq:square_ASD}. We have checked on explicit examples that non-square fishnet integrals $\phi_{m,n}$ with $n\neq m$ are not antipodally self-dual (for any kinematic map).

%%%%%%%%%%%%%%%%%%%%%%%%%%%%%%%%%%%%%%%%%%%%

\section{Proof of antipodal self-duality}
\label{sec:proof}
The proof proceeds in two steps: we first prove~\eqref{flipf}, and then we show that~\eqref{eq:BD} and~\eqref{flipf} together imply~\eqref{eq:square_ASD}.

\paragraph{\underline{Proof of~\eqref{flipf}:}} 
We can easily check that~\eqref{flipf} holds for $p=1$, so we only need to prove it for $p>1$.
A complete proof of~\eqref{flipf} is given in Appendix~\ref{appB}. Here we sketch a simpler argument why~\eqref{flipf} holds, although this argument only allows us to show that~\eqref{flipf} holds modulo terms proportional to $(2\pi i)^2$.

The sum of the terms on the right-hand side of~\eqref{flipf} is defined to be $\ftilde_p$. Our goal is to show that $\Sa(f_p) = \ftilde_p$. We do this by studying the discontinuities of $\Sa(f_p)$ and $\ftilde_p$, as a function of the complex variable $z$.

Since $\Sa(f_p)$ and $\ftilde_p$ are combinations of classical polylogarithms, the only branch points are at $z\in\{0,1\}$. The functions $F(z,\zbar)$ that we are interested in are linear combinations of products of holomorphic and antihomolomorphic functions. The discontinuity around a branch point $z=a$ can be computed as
\beq
\Disc_aF(z,\zbar) = 2\pi i\big[\delta_aF(z,\zbar) - \deltabar_{a}F(z,\zbar)\big]\,,
\eeq
where $\delta_a$ ($\deltabar_a$) computes the discontinuity of $F$ (divided by $2\pi i$) seen as a function of the  variable $z$ ($\zbar$), with the other variable held fixed. 

The function $f_p$ is single-valued, and so the only non-vanishing discontinuities of $\ftilde_p$ come from $L$. Treating $L$ as independent of $f_p$ and differentiating \eqref{flipf}, it is easy to obtain the useful relation
\beq\label{diffLftilde}
\partial_L \ftilde_p = (p-1) \ftilde_{p-1} \,.
\eeq
Since $\delta_0\log z = \deltabar_0\log \zbar =1$, $\delta_1\log z = \deltabar_1\log \zbar = 0$, we find
\beq\bsp\label{eq:disc_f_p_tilde}
\Disc_0\ftilde_p &\,= 2\pi i\,(p-1)\,\log\frac{z}{\zbar}\,\ftilde_{p-1}\,,\\
\Disc_1\ftilde_p &\,= 0\,.
\esp\eeq
We claim that these relations match the discontinuities of $\Sa(f_p)$. First, note that at the level of the symbol, the antipode exchanges discontinuity and derivative operators. Indeed, discontinuities and differentiation remove the first and last entries of the symbol respectively, and the antipode reverses all the symbol letters (up to a sign). The same conclusion holds at function level, up to terms proportional to $(2\pi i)^2$, by noting that discontinuities and derivatives act on the first and last entries, respectively, in the coproduct on MPLs~\cite{Goncharov:2005sla,brownmixedZ,Duhr:2012fh,brownnotesmot}, and the antipode exchanges the two entries of the coproduct.

It follows that in order to understand the discontinuities of $\Sa(f_p)$ up to terms proportional to $(2\pi i)^2$, it is sufficient to study the derivatives of $f_p$. It is easy to see that the ladder integrals have no $(1-z)$ or $(1-\zbar)$ in the final entry of their coproduct; they satisfy
\beq
\lim_{z\to 1}(1-z)\partial_z f_p
=\lim_{\zbar\to 1}(1-\zbar)\partial_{\zbar}f_p=0\,.
\eeq
This translates into the fact that the symbol of $\Disc_1\Sa(f_p) $ vanishes up to terms proportional to $(2\pi i)^2$, in agreement with~\eqref{eq:disc_f_p_tilde}. For the discontinuity at 0, we employ the following identity, whose proof is presented in Appendix~\ref{appA} (see also~\cite{Petkou:2021zhg,Karydas:2023ufs}),
\beq\label{eq:app_proof}
(z\partial_z + \zb \partial_{\zb}) f_p = -(p-1) \log(z\zb) f_{p-1}\,.
\eeq
Using the fact that the antipode exchanges discontinuities and differentiation, we find,
\begin{align}
\big(\delta_0-\deltabar_0)\C S(f_p)&\,\mathrel{\widehat{=}} \C\!\big(\deltabar_0+\delta_0) S(f_p)\\
\nonumber&\,\mathrel{\widehat{=}}
\Sa\Bigl[ \big(\zbar\partial_{\zbar}+z\partial_z)f_p \Bigr]\\
\nonumber&\,\mathrel{\widehat{=}}
(p-1) \log\tfrac{z}{\zbar}\Sa(f_p)\,.
\end{align}
in agreement with~\eqref{eq:disc_f_p_tilde}, and where we use the notation $a\mathrel{\widehat{=}}b$ to indicate that $a$ is equal to $b$ up to terms proportional to $(2\pi i)^2$.

It remains to show that 
$\Sa(f_p) - \ftilde_p \mathrel{\widehat{=}} 0$.
From the previous discussion it follows that $\Sa(f_p) - \ftilde_p$ is single-valued up to terms proportional to $(2\pi i)^2$. Single-valued MPLs are dictated by their holomorphic part.  So it is sufficient to consider only the holomorphic part of~\eqref{flipf}, which means dropping all the terms with $k>0$, as well as the antiholomorphic part of both sides.  But the holomorphic part of $\Sa(f_p)$ is the antipode of the antiholomorphic part of $f_p$.  However, the way single-valued MPLs are constructed (cf.~\cite{BrownSVHPLs,brownSV,Dixon:2012yy,Brown:2013gia,DelDuca:2016lad}), the purely antiholomorphic part is given by the antipode of the holomorphic part, after conjugation.  Thus the second antipode and conjugation return one to the original holomorphic part of $f_p$.  These two terms exactly cancel, and we have shown that $\Sa(f_p) - \ftilde_p \mathrel{\widehat{=}} 0$.

\paragraph{\underline{Proof of~\eqref{eq:square_ASD}:}} We now show that~\eqref{eq:BD} and~\eqref{flipf} together imply~\eqref{eq:square_ASD}. The proof is purely combinatorial, and does not rely on properties of polylogarithms. We can therefore treat $f_1,\ldots,f_p$ and $L$ as independent variables. Let $\Phi$ be the matrix with entries 
$\Phi_{ij} = \tf_{i+j-1}$, $1\le i,j\le m$, so that
$\Sa(\phi_m) = \det\Phi$.  Suppose we can show that $\partial_L \det\Phi=0$. Then we can set $L\to0$ in $\det\Phi$, which sets $\ftilde_p \to f_p$ everywhere, making $\det\Phi$ manifestly equal to $\phi_m$ and so we have proven \eqref{eq:square_ASD}.

In order to show that $\partial_L \det\Phi=0$, we compute the derivative row-by-row, using \eqref{diffLftilde}:
\beq\label{eq:eq_for_proof_1}
\partial_L\det\Phi = \sum_{k=1}^m \det\Phi^{(k)}\,,
\eeq
where $\Phi^{(k)}$ is the matrix with entries
\beq\bsp
\Phi^{(k)}_{ij} &\,= \left\{\begin{array}{ll}
\Phi_{ij} \,,& i\neq k\,,\\
\partial_L\Phi_{kj} \,,& i=k\,,
\end{array}\right.\\
&\,= \left\{\begin{array}{ll}
 \tf_{i+j-1}\,,& i\neq k\,,\\
 (k+j-2)\,\tf_{k+j-2}\,,& i=k\,.
\end{array}\right.
\esp\eeq
Consider now the matrix $\widehat{\Phi}^{(k)}$ defined by subtracting from the $k^{\textrm{th}}$ row of $\Phi^{(k)}$ its $(k-1)^{\textrm{th}}$ row multiplied by $(k-1)$. We find
\beq\bsp
\widehat{\Phi}^{(k)}_{ij}
&\,= \left\{\begin{array}{ll}
 \tf_{i+j-1}\,, & i\neq k\,,\\
 (j-1)\,\tf_{k+j-2}\,, & i= k\,.
\end{array}\right.
\esp\eeq
Since this operation leaves the determinant unchanged, we obtain from~\eqref{eq:eq_for_proof_1}
\beq
\partial_L\det\Phi = \sum_{k=1}^m \det\widehat\Phi^{(k)}\,.
\eeq

For any matrix $M$, we denote by $M[a,b]$ the minor of $M$ obtained by deleting the $a^{\textrm{th}}$ row and the $b^{\textrm{th}}$ column. It is easy to see that 
\beq
\widehat\Phi^{(k)}[k,l] = \Phi^{(k)}[k,l] =\Phi[k,l] \,,
\eeq
because $\widehat\Phi^{(k)}$, $\Phi^{(k)}$ and $\Phi$ only differ in the $k^{\textrm{th}}$ row. If we expand $\det\widehat\Phi^{(k)}$ with respect to its $k^{\textrm{th}}$ row, we find
\beq\bsp\label{eq:detformula}
\partial_L\det\Phi
&\,= \sum_{l=2}^m (l-1)\sum_{k=1}^m (-1)^{k+l}\,\tf_{k+l-2}\,\Phi[k,l]\,.
\esp\eeq

Let us define the $m\times m$ matrix
\beq \label{eq:PsiDef}
\Psi^{(l)}_{ij} = \left\{\begin{array}{ll}
\tf_{i+j-1}\,,& j\neq l\,,\\
\tf_{i+l-2}\,,& j=l\,.
\end{array}\right.
\eeq
It is easy to see that $\Psi^{(l)}[k,l] = \Phi[k,l]$, and so, if we expand the determinant of $\Psi^{(l)}$ with respect to its $l^{\textrm{th}}$ column, we have
\beq
\det\Psi^{(l)} = \sum_{k=1}^m 
(-1)^{k+l} \tf_{k+l-2}\,\Phi[k,l]\,.
\eeq
Comparing this to eq.~\eqref{eq:detformula}, we see that
\beq
\partial_L\det\Phi = \sum_{l=2}^m 
(l-1)\det\Psi^{(l)}\,.
\eeq
But $\det\Psi^{(l)}=0$ for all $2\le l\le m$, because $\Psi^{(l)}$ always contains two identical columns, $j=l-1$ and $j=l$ in \eqref{eq:PsiDef}.  Hence, $\partial_L\det\Phi =0$, which finishes the proof.

%%%%%%%%%%%%%%%%%%%%%%%%%%%%%%%%%%%

\section{Discussion}
\label{sec:discussion}

In this letter we have shown that square fishnet integrals $\phi_m$ are antipodally self-dual, by which we mean that they are invariant under the action of the map $\Sa = \C S$, where $S$ is the antipode and $\C$ is defined in~\eqref{eq:C_def}. So far, antipodal self-duality has only been observed for the MHV four-particle form factor for the chiral part of the stress-tensor supermultiplet~\cite{Dixon:2022xqh}. (A weaker self-symmetry holds for the parity-even part of two-loop MHV amplitudes in planar $\cN=4$ SYM~\cite{Liu:2022vck}.) Our results extend this list to square fishnet integrals, which compute correlators in fishnet theory~\cite{Gurdogan:2015csr}. The structure of our antipodal map $\Sa$ is similar to the previous ones~\cite{Dixon:2022xqh,Liu:2022vck}, and consists of the antipode $S$ on polylogarithms followed by a kinematic map $\C$ that acts on the symbol letters. Importantly, we can formulate the square-fishnet symmetry at the function level for any $m$. Furthermore, we have provided a rigorous proof that holds for any such square fishnet graph.  Also, fishnet graphs have alternate representations related to integrability~\cite{Basso:2017jwq,Derkachov:2019tzo,Derkachov:2020zvv,Basso:2021omx}.  It may be that antipodal self-duality can be understood at a more fundamental level using these representations.

Many conformally-invariant four-point functions can be evaluated in terms of the same class of functions as the fishnet integrals, namely single-valued (harmonic) polylogarithms.  An important class is large $R$-charge correlators, where other determinants of ladder integrals appear~\cite{Coronado:2018cxj,Caron-Huot:2021usw}.
Our results immediately beg the question of whether other classes of correlators or Feynman integrals enjoy an antipodal symmetry. 
We have already mentioned that fishnet integrals $\phi_{m,n}$ with $n\neq m$ are not antipodally self-dual, which can be established independently of the kinematic map for small $m,n$ by counting dimensions of spaces of coproducts.  Also, the generalized fishnet integrals of~\cite{Coronado:2018cxj,Caron-Huot:2021usw} that we inspected do not possess antipodal self-duality for the specific map in $\Sa$. 

We also asked whether other polynomials in the ladder integrals $f_p$ could be antipodally self-dual under $\Sa$, and also satisfy the Steinmann relations~\cite{Steinmann,Steinmann2,Cahill:1973qp}, which imply that the first two entries in the symbol (or their complex conjugates) cannot both be $1-z$. (The square and rectangular fishnets and the integrals of~\cite{Coronado:2018cxj,Caron-Huot:2021usw} all satisfy the Steinmann relations.) We found that these conditions are very restrictive, and there are very few such polynomials at small weights. In particular, there is a unique infinite family of polynomials of degree 2 of this type, given by
\beq\bsp
J_m &\,=2\,f_1\, f_{2m - 1} -(-1)^{m} \, \binom{2m - 2}{m - 1} \, f_{m}^2 \\
&\ \ -2 \sum_{k=2}^{m-1} (-1)^{k}\binom{2m - 2}{k - 1} f_{k} f_{2m - k}\,,
\esp\eeq
where $m\ge 2$ is an integer and the ``loop order'' is $2m$. Note that $J_2= 2\,\phi_2$ is the four-loop $2\times 2$ square fishnet graph. 

The first cubic polynomial in ladder integrals with these properties is the 9-loop fishnet graph $\phi_3$, and at 11 loops we have the combination
\beq\bsp
2\, f_1\, f_5^2&-f_7\, f_2^2-f_2\,f_4\, f_5 +3\, f_2\,f_3\, f_6+2\, f_3\, f_4^2\\
&-3\, f_3^2\, f_5-3\, f_1\, f_4\, f_6+f_1\, f_3\, f_7\,.
\esp\eeq
Understanding whether these polynomials in ladder integrals also compute interesting physical quantities may give a path to uncovering more instances of antipodal self-dualities in QFT.

%%%%%%%%%%%%%%%%%%%%%%%%%%%%%%%%%%%%%%

\begin{acknowledgments}
\emph{Acknowledgments:}
We are grateful to Benjamin Basso, Florian~Loebbert, Anthony~Morales, Didina Serban, and Sven~Stawinski for useful discussions.
We acknowledge the support of the Bethe Center for Theoretical Physics in Bonn during the workshop \emph{Fishnets: Conformal Field Theories and Feynman Graphs} in September 2024, where the ideas presented here were first discussed.
This research was supported by the US Department of Energy under contract
DE--AC02--76SF00515, and by the European Research Council (ERC) under the European Union’s research and innovation programme grant agreement 101043686 (ERC Consolidator Grant LoCoMotive).
Views and opinions expressed
are however those of the author(s) only and do not necessarily reflect those of the
European Union or the European Research Council. Neither the European Union
nor the granting authority can be held responsible for them.

\end{acknowledgments}

%%%%%%%%%%%%%%%%%%%%%%%%%%%%%%%%%%%
%\bibliographystyle{bibliostyle}
%\bibliography{fishnet}

\providecommand{\href}[2]{#2}\begingroup\raggedright\endgroup

%%%%%%%%%%%%%%%%%%%%%%%%%%%%%%%%%%%
\onecolumngrid
%\newpage
\appendix*
\allowdisplaybreaks
\renewcommand{\thesubsection}{\Alph{subsection}}
\renewcommand{\theequation}{\Alph{subsection}.\arabic{equation}}
\setcounter{secnumdepth}{2}

\newpage

\section*{APPENDIX}

\subsection{Proof of~\eqref{eq:app_proof}}
\label{appA}
\setcounter{equation}{0}

We present here the straightforward derivation of~\eqref{eq:app_proof}:

\begin{align}
(z\partial_z + \zb \partial_{\zb}) f_p &=\sum_{j=p}^{2p} \frac{(p-1)! j!}{(j-p)! (2p-j)!}
\biggl[ \cL^{2p-j} \Li_{j-1}(z)
-2(2p-j) \cL^{2p-j-1} \Li_j(z) - (z\leftrightarrow \zb) \biggr]
\nonumber\\
&= \sum_{j=p-1}^{2p-1} \frac{(p-1)! (j+1)!}{(j+1-p)! (2p-j-1)!}
 \cL^{2p-j-1} \Li_j(z)
- \! 2 \! \sum_{j=p-1}^{2p-1} \frac{(p-1)! j!}{(j-p)! (2p-j-1)!} \cL^{2p-j-1} \Li_j(z)-(z\leftrightarrow \zbar)\\
\nonumber&= \sum_{j=p-1}^{2p-2} \frac{(p-1)! j!}{(j-p+1)! (2p-j-2)!}
\cL^{2p-j-1} (\Li_j(z) - \Li_j(\zb)) 
\\
\nonumber&= -(p-1) \log(z\zb) \sum_{j=p-1}^{2(p-1)} \frac{(p-2)! j!}{(j-(p-1))! (2(p-1)-j)!}
\cL^{2(p-1)-j} (\Li_j(z) - \Li_j(\zb)) 
\nonumber\\
\nonumber&= -(p-1) \log(z\zb) f_{p-1}\,.
\end{align}
In the third step we used $(j+1) - 2(j-p+1) = 2p-j-1$.
Eq.~\eqref{eq:app_proof} has also appeared in the context of the expectation value of the charge operator for a free massive boson at finite temperature with a chemical potential~\cite{Petkou:2021zhg,Karydas:2023ufs}.

%%%%%%%%%%%%%%%%%%%%%%%%%%%%%%%%%%%%%%%%
\subsection{Proof of~\eqref{flipf}}
\label{appB}
\setcounter{equation}{0}

Here we provide a purely combinatorial proof of~\eqref{flipf}. In contrast to the proof in the main text, this proof retains all $(2\pi i)^2$ contributions. We first show that we can reduce the proof of~\eqref{flipf} to proving an equality between two homogeneous polynomials. Then we show that the two polynomials have the same coefficients.

\paragraph{\underline{Reduction to an equality between polynomials.}}
We start by working out $\Sa(f_p)$ explicitly. We act with $S$ and $\C$ on~\eqref{eq:ladder} and insert~\eqref{eq:Li_antipode}. We find
\beq
\Sa(f_p) = \sum_{k=p}^{2p}\sum_{l=1}^{k}\frac{(p-1)!k!}{(k-p)!(2p-k)!(k-l)!}\,(-1)^{l}\Li_l(z)\,\log^{k-l} z\,\log^{2p-k}\frac{\zbar}{z}   - (z\leftrightarrow \zbar)\,.
\eeq
If we rearrange the sums according to
\beq
\sum_{k=p}^{2p}\sum_{l=1}^{k} a_{k,l}  = \sum_{l=1}^{2p}\sum_{k=l}^{2p} a_{k,l}=\sum_{l=1}^{2p}\sum_{k=0}^{2p-l} a_{2p-k,l}\,,
\eeq
we arrive at
\beq\label{eq:LHS_reduced}
\Sa(f_p) = \sum_{l=1}^{2p}(p-1)!\,(-1)^l\,\Li_l(z)\,P^{(1)}_{p,l}(\log z,\log\zbar)   - (z\leftrightarrow \zbar) \,,
\eeq
where we defined the polynomial
\beq
P^{(1)}_{p,l}(x,y) = \sum_{k=0}^{2p-l}\frac{(2 p - k)!}{(p - k)!k!(2 p - k - l)!} \,x^{2 p - k - l} (y-x)^{k}\,.
\eeq
Note that $P^{(1)}_{p,l}(x,y)$ is a homogeneous polynomial of degree $2p-l$. We write
\beq
P^{(1)}_{p,l}(x,y) = \sum_{r=0}^{2p-l}c_{p,l,r}\,x^{2p-l-r}\,y^r\,.
\eeq

We now perform similar manipulations on the right-hand side of~\eqref{flipf}. We first let $k\to p-k$ in \eqref{flipf}, and then rearrange the sums according to
\beq
\sum_{k=1}^{p}\sum_{l=k}^{2k} a_{k,l}  = \sum_{l=1}^{2p}\sum_{k=1}^{\min(l,p)} a_{k,l}=\sum_{l=1}^{2p}\sum_{k=1}^{l} a_{k,l}\,.
\eeq
In the last step we replaced the upper summation limit $\min(l,p)$ by $l$. This is justified in our case, because the summand contains a factor $\tfrac{1}{(p-k)!}$, which vanishes for $k>p$. We then find
\begin{align}\label{eq:RHS_reduced}
&\sum_{k=0}^{p-1}\binom{p-1}{k}\,\log^k z\,\log^k\zbar\,f_{p-k} = 
\sum_{k=1}^{p}\binom{p-1}{p-k}\,\log^{p-k} z\,\log^{p-k}\zbar\,f_k \nonumber\\
&= \sum_{l=1}^{2p}(p-1)!\,(-1)^l\,\Li_l(z)\,P^{(2)}_{p,l}(\log z,\log\zbar) - (z\leftrightarrow \zbar)\,,
\end{align}
with
\beq
P^{(2)}_{p,l}(x,y) = \sum_{k=1}^l \frac{l!}{(l - k)!(2 k - l)!(p - k)!}\, x^{p - k} \,y^{p - k}\,(x+y)^{2 k - l}\,.
\eeq
It is easy to see that $P^{(2)}_{p,l}(x,y)$ is again homogeneous of degree $2p-l$, and we write
\beq
P^{(2)}_{p,l}(x,y) = \sum_{r=0}^{2p-l}d_{p,l,r}\,x^{2p-l-r}\,y^r\,.
\eeq

Comparing~\eqref{eq:LHS_reduced} and~\eqref{eq:RHS_reduced}, we see that~\eqref{flipf} holds if $P^{(1)}_{p,l}(x,y) = P^{(2)}_{p,l}(x,y)$, i.e., if they have the same coefficients,
$c_{p,l,r} = d_{p,l,r}$, for $1\le l\le 2p$ and $0\le r\le 2p-l$.  Next we compute these two sets of coefficients.

\paragraph{\underline{Computation of the coefficients $c_{p,l,r}$.}}
Since $P^{(1)}(x,y)$ is homogeneous of degree $2p-l$, its coefficients are given by
\beq
c_{p,l,r} = \frac{1}{r!}\partial_{y}^rP^{(1)}_{p,l}(x,y)|_{x=1,y=0} = \sum_{k=r}^{2p-l}\frac{(2 p - k)!}{(p - k)!(2 p - k - l)!r!(k - r)!} (-1)^{k - r} \,.
\eeq
In order to form this sum, we consider the regulated version
\beq
C_{p,l,r}(\eps) = \sum_{k=r}^{2p-l}\frac{\Gamma(1+\eps+2 p - k)}{\Gamma(1+2\eps+p - k)(2 p - k - l)!r!(k - r)!} (-1)^{k - r} \,.
\eeq
Using $\Gamma(1+n) = n!$ for positive integer $n$, it is easy to see that 
\beq\label{eq:Clim}
C_{p,l,r}(\eps=0) = c_{p,l,r}\,.
\eeq
This regularization will be needed in order to perform the sum. Writing the factorials and the $\Gamma$ functions in terms of Pochhammer symbols,
\beq
(a)_{\pm k} = \frac{\Gamma(a\pm k)}{\Gamma(a)}\,,
\eeq 
and using the identity
\beq
(a)_{-k} = \frac{(-1)^k}{(1-a)_k}\,,
\eeq
we see that $C_{p,l,r}(\eps)$ can be expressed in terms of Gauss' hypergeometric function
${}_2F_1(a,b;c;z) = \sum_{n=0}^\infty\frac{(a)_n(b)_n}{(c)_n}\,\frac{z^n}{n!}$.
We find
\beq\bsp
C_{p,l,r}(\eps) &\,= \frac{\Gamma (1+2 p-r+\epsilon ) }{\Gamma (1+r) \Gamma (1-l+2 p-r) \Gamma (1+p-r+2 \epsilon )}\,{} _2F_1(l-2 p+r,r-p-2 \epsilon ;r-2 p-\epsilon ;1)\\
&\, = \frac{\Gamma (r-2 p-\epsilon ) \Gamma (1+2 p-r+\epsilon ) \Gamma (p-l-r+\epsilon )}{\Gamma (1+r) \Gamma (-l-\epsilon ) \Gamma (\epsilon -p) \Gamma (1-l+2 p-r) \Gamma (1+p-r+2 \epsilon )}\,,
\esp\eeq
where in the last step we used the identity ${}_2F_1(a,b;c;1) = \frac{\Gamma(c)\Gamma(c-a-b)}{\Gamma(c-a)\Gamma(c-b)}$.
The final expression makes manifest the need to work with the regularized version: for $\eps\neq0$ the factors $\Gamma(-l-\eps)$ and $\Gamma(\eps-p)$ in the denominator are divergent for all positive integer values of $p$ and $l$. Nonetheless, from~\eqref{eq:Clim} we know that the limit $\eps\to0$ must be smooth. 

We now compute this limit. The value of the limit depends crucially on the values of the integer parameters $p$, $l$ and $r$, since those determine which $\Gamma$ functions are divergent. The parameters $(p,l,r)$ take values in the range $1\le l\le 2p$ and $0\le r\le 2p-l$. It is easy to see that for all values of $(p,l,r)$ in that range, we have
\beq\bsp
\Gamma(-l-\eps), \Gamma(\eps-p), \Gamma (r-2 p-\epsilon ) &\,=\ord\big(\eps^{-1}\big)\,, \\
\Gamma (1+2 p-r+\epsilon ) &\, =\ord\big(\eps^0\big)\,.
\esp\eeq
For the remaining two $\eps$-dependent $\Gamma$ functions, we need to distinguish cases. We start by discussing the case $p\le l$. In that case we have $\Gamma(p-l-r+\eps) = \ord\big(\eps^{-1}\big)$.
The behavior of the other $\Gamma$ function depends on the value of $r$. 
\begin{itemize}
\item If $p\le r\le 2p-l$, we have $\Gamma(1+p-r+2\eps) = \ord\big(\eps^{-1}\big)$, and we find
$C_{p,l,r}(\eps) =  \ord(\eps)$.
\item If $0\le r\le p$, we have $\Gamma(1+p-r+2\eps) = \ord\big(\eps^{0}\big)$, and if we expand all $\Gamma$ functions to their leading order, we find,
\beq\label{eq:coeff_final}
C_{p,l,r}(\eps) = \frac{l!\, p!}{r! (p-r)! (2p-l-r)! (l-p+r)!} + \ord(\eps)\,.
\eeq
\end{itemize}
 Let us now turn to the case $l < p$. If $p-r-l$ and $p-r+1$ have the same sign, we find $C_{p,l,r}(\eps) =  \ord(\eps)$. Otherwise, we must necessarily have $p-r-l\le 0\le p-r+1$. Expanding all $\Gamma$ functions to leading order, we find the same expression as in~\eqref{eq:coeff_final}.

Finally, we note that~\eqref{eq:coeff_final} is valid for all values of $(p,l,r)$ in the range $1\le l\le 2p$ and $r\le 2p-l$. Indeed, in those cases where $C_{p,l,r}(\eps) =  \ord(\eps)$, the expression in~\eqref{eq:coeff_final} contains a divergent factorial in the denominator. Hence, we conclude that in all cases
\beq\label{eq:cplr}
c_{p,l,r} = \frac{l!\, p!}{r! (p-r)! (2p-l-r)! (l-p+r)!}\,,\qquad 1\le l\le 2p\,,\qquad 0\le r\le 2p-l\,.
\eeq

\paragraph{\underline{Computation of the coefficients $d_{p,l,r}$.}}
Since $P^{(2)}(x,y)$ is homogeneous of degree $2p-l$, its coefficients are given by
\beq\bsp
d_{p,l,r} &\,= \frac{1}{r!}\partial_{y}^rP^{(2)}_{p,l}(x,y)|_{x=1,y=0} = \sum_{k=p-r}^{l}\frac{l! }{(2 k-l)! (l-k)! (p-k)!}\,\binom{2 k-l}{k-p+r}\\
&\, = \sum_{k=0}^{l+p-r}\frac{l!}{(k-l)! (2p-k-r)! (k-2 p+2 r)! (p+l-k-r)!} \,.
\esp\eeq
This sum can be evaluated using the generalization of the Vandermonde identity from~\cite{vandermonde}:
\beq\label{eq:vandermonde}
\sum_{K=0}^R\frac{K!}{(K-L)!}\,\binom{M}{K}\binom{N}{R-K} =\frac{M!}{(M-L)!}\,\binom{M+N-L}{R-L}\,.
\eeq
For $L=0$ we recover the classical Vandermonde identity. We can apply~\eqref{eq:vandermonde} with $(L,M,N,R) = (l,2p-r,r+l-p,l+p-r)$. We find
\beq
d_{p,l,r} =  \frac{l!\, p!}{r! (p-r)! (2p-l-r)! (l-p+r)!} = c_{p,l,r}\,,\qquad 1\le l\le 2p\,,\qquad 0\le r\le 2p-l\,.
\eeq
We thus see that $P^{(1)}_{p,l}(x,y)$ and $P^{(2)}_{p,l}(x,y)$ are equal, and so~\eqref{flipf} holds.

\end{document}